%% file: main.tex
\documentclass[a4paper,UKenglish,cleveref,autoref,thm-restate]{lipics-v2021}
\usepackage{multirow}
\usepackage{graphicx}
\usepackage{textcomp}
\usepackage{tikz}
\usepackage{pgfplots}
\usepackage{subfiles}
\usepackage{listings}
\usepackage{xcolor}
\usepackage{subcaption}
\usepackage{helvet}
\usepackage{todonotes}
\usepackage{caption}

\definecolor{codegreen}{rgb}{0,0.6,0}
\definecolor{codegray}{rgb}{0.5,0.5,0.5}
\definecolor{codepurple}{rgb}{0.58,0,0.82}
\definecolor{backcolour}{rgb}{0.95,0.95,0.92}

\lstdefinestyle{mystyle}{               
    basicstyle=\ttfamily\scriptsize,
    keywordstyle=\color{magenta}
}

\AtBeginDocument{%
  }

\title{WasmWalker: Path-based Code Representations for Improved WebAssembly Program Analysis}

\titlerunning{WasmWalker: Path-based Code Representations for Improved WebAssembly\dots} 

\author{Mohammad Robati Shirzad}{University of Waterloo, Waterloo, Ontario, Canada}{mrobatis@uwaterloo.ca}{https://orcid.org/0000-0002-6297-8229}{}

\author{Patrick Lam}{University of Waterloo, Waterloo, Ontario, Canada}{patrick.lam@uwaterloo.ca}{https://orcid.org/0000-0001-8278-5400}{}

\authorrunning{M. Robati Shirzad and P. Lam} 

\Copyright{Jane Open Access and Joan R. Public} 



\category{} 

\relatedversion{} 

\nolinenumbers 

\begin{document}

\maketitle

\begin{abstract}
    WebAssembly, or Wasm, is a low-level binary language that enables execution of near-native-performance code in web browsers. Wasm has proven to be useful in applications including gaming, audio and video processing, and cloud computing, providing a high-performance, low-overhead alternative to JavaScript in web development. The fast and widespread adoption of WebAssembly by all major browsers has created an opportunity for analysis tools that support this new technology. \\
    Deep learning program analysis models can greatly benefit from the program structure information included in Abstract Syntax Tree (AST)-aware code representations. To obtain such code representations, we performed an empirical analysis on the AST paths in the WebAssembly Text format of a large dataset of WebAssembly binary files compiled from source packages in the Ubuntu 18.04 repositories. After refining the collected paths, we discovered that only 3,352 unique paths appeared across all of these binary files. \\
    With this insight, we propose two novel code representations for WebAssembly binaries. These novel representations serve not only to generate fixed-size code embeddings but also to supply additional information to sequence-to-sequence models. Ultimately, our approach helps program analysis models uncover new properties from Wasm binaries, expanding our understanding of their potential. We evaluated our new code representation on two applications: (i) method name prediction and (ii) recovering precise return types. Our results demonstrate the superiority of our novel technique over previous methods. More specifically, our new method resulted in 5.36\% (11.31\%) improvement in Top-1 (Top-5) accuracy in method name prediction and 8.02\% (7.92\%) improvement in recovering precise return types, compared to the previous state-of-the-art technique, \textsc{SnowWhite}.  
\end{abstract}

\section{Introduction}
\subfile{introduction}

\section{\label{sec:motivating}Motivating Example}
\subfile{motivating}

\section{\label{sec:methodology}Methodology}
\subfile{methodology}

\section{\label{sec:evaluation}Evaluation}
\subfile{evaluation}

\section{\label{sec:discussion}Discussion}
\subfile{discussion}

\section{\label{sec:related}Related Work}
\subfile{related}

\section{\label{sec:conclusion}Conclusion}
\subfile{conclusion}

\bibliography{refs.bib}

\end{document}

%% file: introduction.tex
WebAssembly (Wasm) is a cross-platform low-level binary format that enables browsers to execute code with near-native performance. Wasm is an open standard, currently maintained by the World Wide Web Consortium (W3C). The technology was created as a collaboration between W3C, Mozilla, Google, Microsoft, and Apple~\cite{haas2017bringing}. The development of Wasm began in 2015, and version 1.0 was officially released in 2017. Wasm seeks to go beyond the limitations of JavaScript, providing a secure platform for executing programs in web browsers. Some of JavaScript's problems include: 1) its dynamic nature requires heroic efforts to execute efficiently; 2) as a client-side language, JavaScript has limited access to memory and the file system; and 3) JavaScript's multiprocessing capabilities are restricted. Wasm was developed with the aim of filling these gaps. In terms of facilitating the deployment of platform-independent code and providing a security model that ensures that the system functions in a secure and reliable manner, the role of Wasm on the client is comparable to the role that Java performed in the early 2000s on the server with the Java Virtual Machine.

Wasm is becoming more and more popular: Wasm's adoption has created new opportunities for web development, enabling programmers to consider using Wasm in performance-critical applications, such as cloud computing, video conferencing, and gaming. Some well-known programming languages, including C, C++, and Rust, now have Wasm compilers available. Wasm's popularity serves as an invitation for more tools and techniques to be created to support this newly-emerged technology.

This context motivated us to develop new tools and techniques for static analysis of Wasm. We focus particularly on techniques that can help with deep learning: the application of deep learning in static analysis has brought about significant advancements in the field. Deep learning models are able to analyze large datasets quickly and can predict properties of unseen data with high accuracy and consistency. This has made deep learning one of the most popular tools for static analysis, and its popularity is growing as the learning methods are constantly evolving. Deep learning can be used for many interesting properties that we target in static analysis. For instance, researchers have developed deep learning tools to detect security vulnerabilities in high level programming languages~\cite{li2018vuldeepecker, zou2019mu, batur2021novel}, and also native binaries~\cite{aumpansub2022learning}. Other applications include defect prediction~\cite{manjula2019deep, qiao2020deep} and clone detection~\cite{li2017cclearner, wang2020detecting}. 

The input provided to deep learning models greatly impacts their performance and accuracy. If the input contains the right information and is \emph{structured} correctly, models can make highly-accurate predictions. However, if the input data contains redundant information or is poorly structured, this can slow down the training process and reduce the accuracy of the models. Our primary goal is to devise a useful structure for providing information about code as input to models.

Model inputs can be fixed-sized or variable-sized. Fixed-sized inputs work when the data's number of features is known: datapoints are vectors of numbers. Vector dimensions are associated with defined features. In static analysis, inputs are usually programs or parts of programs. Yet, computer programs are inherently variable-sized information. Thus, to use them as inputs to, for example, feedforward neural networks, we must embed them into fixed-sized vectors---a process known as code embedding~\cite{chen2019literature}. Researchers have proposed many approaches for embedding high-level code~\cite{allamanis2015suggesting, devlin2017semantic, buch2019learning, alon2019code2vec} and native binaries~\cite{ben2018neural, zuo2018neural, redmond2018cross, xu2017neural}.

Unlike fixed-sized inputs, variable-sized inputs do not have an a priori known number of features. Variable-sized inputs are typically used when features are more complex and cannot be easily transformed into numerical values. In the case of static analysis, the input can directly be the sequence of program statements. More complex (and hence more expensive) deep learning models, including RNNs, LSTMs~\cite{hochreiter1997long}, Transformer Networks~\cite{vaswani2017attention}, and GPT-3~\cite{brown2020language}, process variable-sized inputs. 

In this project, we aim to make information extracted from Wasm ASTs available as input to deep learning models. Towards this goal, we introduce WasmWalker, a pipeline for extracting AST information from Wasm binaries and representing that information as both fixed-sized and variable-sized model inputs. Using WasmWalker, we conducted an empirical study to find the most frequent paths that emerge in the abstract syntax tree of WebAssembly Text (WAT) format of Wasm binaries across a large dataset. We used, as a primary dataset, the same one used in \textsc{SnowWhite}~\cite{lehmann2022finding}, a study that offers a framework for recovering precise C/C++ high level types from low-level limited Wasm type system. The dataset includes 6.3 million type-labeled Wasm samples, extracted from 300,905 object files, which were compiled from 4,081 C and C++ Ubuntu packages. We also augmented the C/C++ dataset with a secondary set of 10 mid-to-large Rust projects to cross-check the generalizability of our paths set. We emphasize, however, that the representations that we present in this work are novel; while \textsc{SnowWhite} uses only a naive representation that consists of taking the last 20 instructions of a method, we propose and evaluate easy-to-compute representations that work better than \textsc{SnowWhite}'s naive representation. We emphasize that our novel code representations can be used in models that use either fixed-size or variable-sized inputs, versus \textsc{SnowWhite}, which can only be used in models which accept variable-sized inputs.

Initially, WasmWalker gathered over 800,000 root-to-leaf paths. We refined these paths to obtain a more compact and abstract representation that respects the nested structure of ASTs and the order of non-terminals within root-to-leaf paths. Our refined paths include 3,352 paths encoding conditional and loop nested structures. Building on the refined paths, we developed a fixed-sized feature vector for each Wasm function---the path vector. Each dimension of our path vectors corresponds to one of the 3,352 paths we identified, and the numerical value for that dimension represents the number of times that path occurred in the Wasm function's AST. This construction of our path vectors yields some interpretability---there is a reason that each value in a path vector has a particular value.


Using our path vectors, we developed two different code representations for Wasm functions: (1) a (variable-sized) path sequence, that indicates the appearance of each path along with a numerical value that corresponds to the number of times that path appeares in the Wasm function, and (2) a (fixed-size) code embedding, similar to code2vec~\cite{alon2019code2vec}, that reduces the dimensionality of our path vectors from 3,352 to 50.

Although our path-based approach, which can produce code embeddings, is more broadly applicable than that of \textsc{SnowWhite}~\cite{lehmann2022finding}, we also carried out a head-to-head test showing improvements over that previous work. Thus, we evaluated the prediction accuracy of our code representations using the dataset provided in \textsc{SnowWhite}. We conducted two experiments. First, we attempted to predict actual method names based on Wasm binaries. Second, similar to \textsc{SnowWhite}, we aimed to recover high-level C/C++ return types from Wasm primitive types. Our findings indicate that a hybrid approach, where a selection of instructions is concatenated to our path sequence, resulted in the best models. Specifically, our new representation led to a 5.36\% (11.31\%) improvement in Top-1 (Top-5) accuracy in method name prediction and a 8.02\% (7.92\%) improvement in recovering precise return types. Also, we demonstrate the effectiveness of our embeddings in clustering method names that contain semantically similar method bodies close together.

\textbf{Contributions} This paper makes the following contributions:
\begin{itemize}
    \item We conduct the first empirical study of the most frequent paths in the ASTs of Wasm binaries over a large-scale dataset.
    \item We propose two novel and easy-to-compute representations of Wasm functions: (1) a path sequence containing AST path information, and (2) a code embedding that leverages the information we gathered about paths. These representations enable the use of Wasm functions as input to deep learning techniques---even those techniques that require code embeddings as inputs. Previous approaches could only feed into variable-size inputs, whereas our representation works for both variable-size and fixed-size inputs.
    \item We compare our approach to previous work on two deep learning use cases: method name prediction and recovering precise return types. Our path sequence in combination with Wasm instructions improves accuracy over the previous state-of-the-art.
\end{itemize}

\newpage
We show that incorporating AST knowledge into the input for models analyzing Wasm binaries enhances their effectiveness. We believe our novel code representations help uncover new properties from Wasm binaries, expanding our understanding of their potential.


\textbf{Data Availability Statement} Our software pipeline, as well as the datasets generated and analyzed in this work, are available at the anonymyzed \url{https://zenodo.org/records/10983638}.

%% file: motivating.tex
In this section, we demonstrate, by example, how the WasmWalker pipeline works. We start with Wasm binaries, generated from C/C++ source code using emscripten. Next, we tranform Wasm binaries to WAT files and extract AST paths from them. Finally, we show how WasmWalker computes two different code representations using the AST paths.

The primary dataset that we used in this work contains Wasm binaries that are created by compiling C/C++ code using the Emscripten toolchain, although Section~\ref{sec:alternative-compilation} describes our evaluation of binaries generated from Rust. Emscripten~\cite{emscripten} is an open source compiler toolchain that compiles C/C++ programs, or any other languages that use LLVM as a part of their compilation process, into WebAssembly. LLVM~\cite{llvm} is a compiler infrastructure that offers a set of modular and reusable compiler technologies. The LLVM Intermediate Representation (LLVM IR) is an assembly-like language that acts as a middle ground between (programmer-written) high-level code and (executable) native code. An intermediate representation makes it possible for compiler developers to only focus on writing the frontend part of the compiler and to delegate the responsibility for dealing with the target architecture to the back-end part of the compilation toolchain. Emscripten's primary work is performed by a series of Wasm-specific optimization passes. To produce LLVM IR code, Emscripten uses Clang, which is a part of the LLVM compiler infrastructure. As a backend, i.e. to transform LLVM IR to Wasm binaries, after using LLVM's optimizer tools to improve performance and efficiency, Emscripten uses the Binaryen tool~\cite{binaryen} to produce Wasm binaries.

Our objective is to analyze the Wasm binary files that we obtain. However, these binary files contain a stream of binary values that make it challenging to perform structural analysis. To overcome this, we use the WebAssembly Text (WAT) format, which is a human-readable representation of Wasm code. We convert a Wasm binary to its respective WAT file using the wasm2wat module of WABT~\cite{wabt}, a binary toolkit for Wasm. While there are some similarities between LLVM code and WAT code, they serve different purposes. LLVM files are an intermediate representation used during compilation, optimization, and code generation and are not intended for human consumption. In contrast, WAT files are commonly used for debugging and inspecting Wasm code, and they offer a tree-like structure of Wasm binaries that can be parsed into an AST for structural analysis, which is crucial for our analysis objectives.

Analogues to our WAT-based approach for other languages should be relatively straightforward to implement. Our technique leverages the fact that a structured AST is available for WAT; such an AST is generally available when compiling from source. Applying our technique to flat IRs or binaries would require first applying decompilation techniques~\cite{cifuentes94:_rever,basque2024ahoy} to obtain a structured AST.
\newpage

\newpage
\begin{figure*}[!t]
    \centering
    \begin{subfigure}[b]{0.45\linewidth}
        \begin{lstlisting}[language=c,basicstyle=\ttfamily\small,backgroundcolor=\color{white}]
int sign(int a) {
  int s = 0;
  if(a > 0) {
    s = 1;
  } else if (a < 0) {
    s = -1;
  }
  return s;
}
        \end{lstlisting}
    \caption{C code}
    \label{fig:codes:c}
    \end{subfigure}
    \begin{subfigure}[b]{0.45\linewidth}
        \begin{lstlisting}[language=llvm,backgroundcolor=\color{white}]
define i32 @sign(i32 %a) #0 {
  entry:
    %a.addr = alloca i32, align 4
    %s = alloca i32, align 4
    store i32 %a, i32* %a.addr, align 4
    ...
  ...
  if.end:
    br label %if.end3
  if.end3:
    %2 = load i32, i32* %s, align 4
    ret i32 %2
}
        \end{lstlisting}
        \caption{LLVM IR code}
        \label{fig:codes:llvm}
    \end{subfigure}
    \begin{subfigure}[b]{0.45\linewidth}
        \begin{lstlisting}[language=llvm,backgroundcolor=\color{white}]
00000000: 0061 736d 0100 0000
00000008: 0192 8180 8000 1660 
00000010: 017f 017f 6000 017f
00000018: 6003 7f7f 7f01 7f60
00000020: 0000 6001 7f00 6003
00000028: 7f7e 7f01 7e60 027f
00000030: 7f01 7f60 067f 7c7f
00000038: 7f7f 7f01 7f60 027f
00000040: 7f00 6002 7e7f 017f 
00000048: 6004 7f7e 7e7f 0060
00000050: 047f 7f7f 7f01 7f60
...
        \end{lstlisting}
        \caption{Wasm binary code}
        \label{fig:codes:wasm}
    \end{subfigure}
    \begin{subfigure}[b]{0.45\linewidth}
        \begin{lstlisting}[backgroundcolor=\color{white}]
(func (type 0) (param i32) (result i32)
  (local i32 ... i32)
  global.get 0
  local.set 1
  ...
  block 
    block 
      local.get 11
      i32.eqz
      br_if 0
      ...
  return)
        \end{lstlisting}
        \caption{WAT code}
        \label{fig:codes:wat}
    \end{subfigure}
    \begin{subfigure}[b]{0.45\linewidth}
        \begin{lstlisting}[basicstyle=\ttfamily\small,backgroundcolor=\color{white}]
func,global.get,0
func,local.set,1
...
func,block,block,local.get,11
func,block,block,i32.eqz
func,block,block,br_if,0
...
func,return
        \end{lstlisting}
        \caption{Extracted raw paths with DFS}
        \label{fig:codes:paths}
    \end{subfigure}
    \begin{subfigure}[b]{0.45\linewidth}
      \fontsize{7}{10}\selectfont
        \begin{flalign*}
&\text{path\_vector} = 
  \begin{array}{lll}
  [1, & 0, & 2, \\
  \:\:0, & ..., & 4]
  \end{array}_{3,352}
 && \\\\
&\text{path\_sequence} = \langle 1,1 \rangle\,\,\langle 3,1 \rangle\,\,...\,\,\langle 3352,2 \rangle && \\\\
&\text{code\_embedding} =
  \begin{array}{lll}
    [0, & 0, & 0.8898107, \\
    \:\:..., & ..., & 0.60188985]
    \end{array}_{50} &&
        \end{flalign*}
      \caption{Code representations}
      \label{fig:codes:representations}
    \end{subfigure}

    \caption{This figure illustrates the files generated while compiling C code to Wasm binaries and then extracting paths from WAT files.}
    \label{fig:codes}
\end{figure*}

Figure~\ref{fig:codes} illustrates different code files that we create until we extract the necessary AST information for further analysis. Subfigure~\ref{fig:codes:c} shows the C source code for a simple \verb+sign()+ function that outputs the sign of its integer input. There are three possible outputs: 0 if the input is 0; and 1 or -1 if the input is a positive or negative integer, respectively. Using Emscripten, we then compile the C function to a LLVM IR function, shown in subfigure~\ref{fig:codes:llvm}. We can select different optimization levels to create LLVM code---the code shown in the subfigure is generated using the -O3 option. At the last stage of compilation, Emscripten generates Wasm binaries, shown in subfigure~\ref{fig:codes:wasm}. As discussed earlier, a Wasm binary is a stream of binary values. To ease analysis, we convert it to WAT format. As illustrated in subfigure~\ref{fig:codes:wat}, the WAT files yield a tree-like intermediate representation of Wasm binaries. We can extract all paths within this tree using a Depth-First Search (subfigure~\ref{fig:codes:paths}). Note that the shown extracted paths are raw paths. We carry out a refinement process to reduce the number of paths, and only keep paths that are more semantically valuable.  Section~\ref{sec:methodology} discusses the refinement process in greater depth.

We extract the AST paths from all the Wasm binaries within our dataset. The total number of all the paths after refinement is 3,352. We represent each Wasm function using a feature vector, which we call the path vector. This vector has 3,352 dimensions. The $i^{\text{th}}$ element of the vector corresponds to the number of times path $i$ occurs in the AST of that function. We build two code representations using our path vectors: (1) a path sequence that is a sequential representation of the path vector, but with normalized occurrence values, and (2) a code embedding similar to that of code2vec~\cite{alon2019code2vec}, which is a distributional representation of Wasm code using fewer dimensions (namely, 50). Subfigure~\ref{fig:codes:representations} shows the structure of our path vector, path sequence, and code embedding. In this subfigure, \emph{path\_vector} has value 2 at index 2. This means that the third path (path vector indices are zero-based) in our paths set appeared twice in our AST. \emph{path\_sequence} yields a sequence representation of \emph{path\_vector}, after removing zero-valued paths and normalizing the values. \emph{code\_embedding} embeds \emph{path\_vector} into a 50-dimensional vector using a feedforward-neural network. Like code2vec~\cite{alon2019code2vec}, creating embeddings requires a target property; in this work, we use method names as the target property, but the process works equally well with other target properties.

In the next section we describe WasmWalker, the pipeline we developed for extracting paths from WAT files. Then we discuss empirical results about extracted paths, the decisions behind our path refinement process, and more details about our code representations.

%% file: methodology.tex
Our objective is to encode information about a complete Wasm function to an interpretable vector. Similar to Feng et al.~\cite{feng2016scalable}, a naive approach could be to build a feature vector for a function, where the features are general metadata about the function (e.g. number of variables). The problem with that approach is that it does not contain any information about code structure. We take a different approach: we exploit information encoded in the AST structure of Wasm programs, and hence what the program does, to form feature vectors.

Wasm binaries are generally being produced by tools that leverage LLVM compiler infrastructure: in addition to supporting WebAssembly, LLVM provides a high-performance and well-documented toolchain for producing WebAssembly binaries.

However, to study the AST form of the Wasm binaries, we need to convert the binaries to WebAssembly Text (WAT) format, which provides a more structured representation of Wasm binaries. We use wasm2wat~\cite{wabt} for that purpose. The output of wasm2wat is a textual representation of the WebAssembly module that closely mirrors the binary format, but with additional annotations and tree-like formatting to make the module structure more clear.

To process the nested structure of WAT files, we need to linearize the paths so that we can study their content. code2vec~\cite{alon2019code2vec} extracts all leaf-to-leaf paths in the AST. That yields a $\Theta(n^2)$ space explosion given $n$ terminals; for further discussion, see Section~\ref{sec:related}. Unlike code2vec, we instead record all root-to-leaf paths for two primary reasons: (1) to avoid the quadratic explosion, and (2) it is not clear that terminals in WAT ASTs contain useful information. The terminals are memory offset values and can easily change with slight modifications to the original program, even when the semantics stay the same. For instance, if in an alternative implementation of a subroutine, we use an extra variable, the memory locations of other variables may shift to make room for the new variable: the compiler allocates memory for variables based on their type and size. This phenomenon is due to working at a lower (IR) level than the source AST, where the terminals are usually variable names.

\newpage
\begin{figure*}[!t]
    \centering
    \includegraphics[scale=0.5]{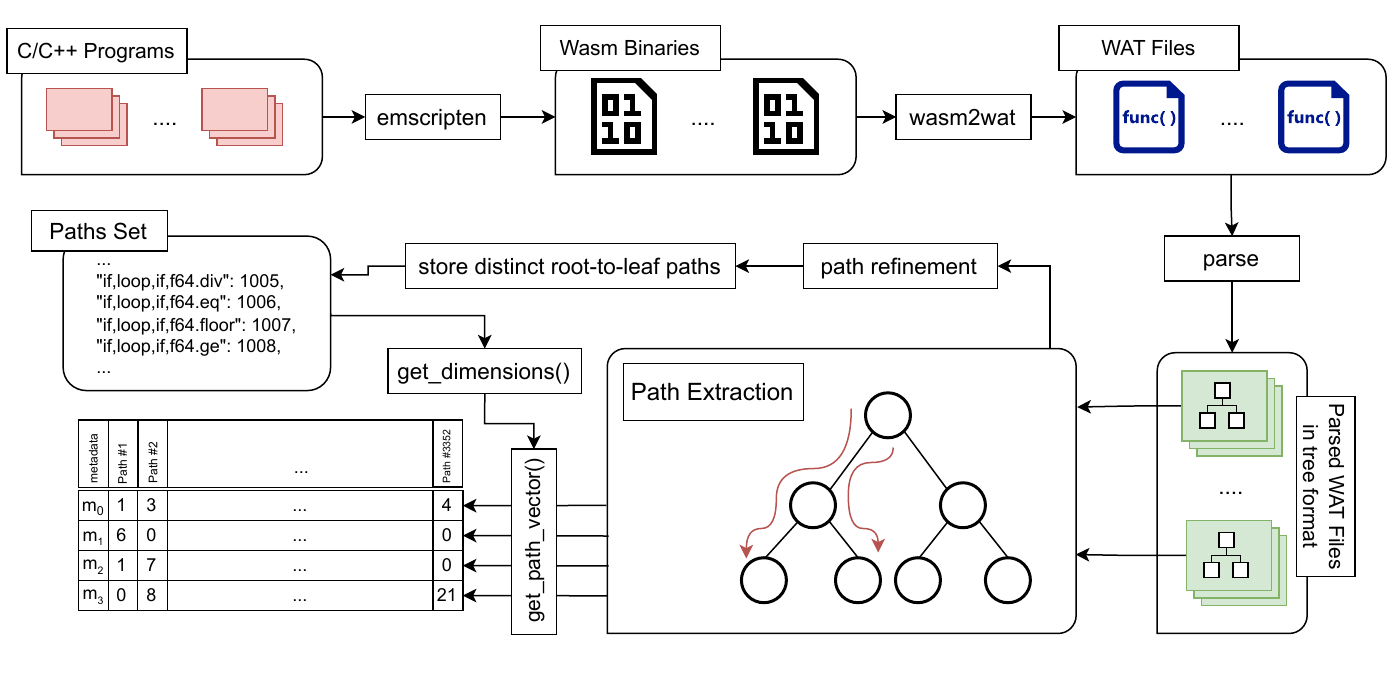}
    \caption{Overview of the WasmWalker pipeline}
    \label{fig:overview}
\end{figure*}
\newpage

Figure~\ref{fig:overview} shows an overview of how we collect the common paths, store them in our paths set, and leverage them to generate path vectors. This figure illustrates the use of emscripten and wasm2wat for transforming C/C++ code to WAT files. Then, we parse the WAT files and extract the raw paths. The raw paths are then refined (we present the refinement process in Section~\ref{sec:collapse}~and~\ref{sec:drop}) and stored in an ordered paths set with unique numbers associated with each path.

Our pipeline, WasmWalker, traverses a WAT file's AST using Depth-First Search and extracts paths inside the subtree of a target function, as the WAT file can contain multiple functions. We refer to internal nodes in our AST as nonterminals and the leaves as terminals. The paths start with the nonterminal ``\verb+func+'' and end with a terminal (see subfigure~\ref{fig:codes:paths}). We drop the keyword ``\verb+func+'', since it is the same for all paths, and also the terminal at the end. The resulting path is a sequence of nonterminals. Note that our use of ``terminal'' and ``nonterminal'' does not match their use in parsing.

WasmWalker is first applied to the training data to populate the paths set. The pipeline is then used again to generate path vectors, which contain the frequency of each path at its corresponding location in the set. The paths set is queried to obtain the index associated with a path. Each path vector also includes metadata, such as the function's name and high-level return type, collected from DWARF debugging information.

\subsection{Common paths within WAT ASTs}
\label{sec:common-paths}

We conducted an empirical study to find the common root-to-leaf paths that appear in WAT ASTs. Our dataset is comprised of training, validation, and test data. To carry out a more accurate evaluation and avoid inflating the accuracy scores of our models, we extracted the common paths only from the training portion of our dataset. During testing, if an unseen path appears, WasmWalker would disregard it. However, such occurrences did not happen; all discovered paths in the test data had previously been seen during training.

We extracted a total number of 807,972 raw paths. This number of paths would pose significant storage challenges due to the sheer volume of dimensions for each data point. Moreover, these vectors would likely be highly sparse, leading to inefficient learning of data. Thus, we reduce it through a path refinement process.

\begin{table}[!htbp]
    \hspace*{-0.7cm}
    \begin{minipage}[t]{0.55\textwidth}
        \caption{Top most common root-to-leaf paths}
        \label{table:path_count} 
        \begin{tabular}{ r l r r } 
            \hline
            Rank & Path & \# & \% \\
            \hline
            1 & \verb+local.get+ & 31,886,923 & 19.67 \\
            2 & \verb+loop,local.get+ & 17,887,693 & 11.03\\
            3 & \verb+local.set+ & 17,397,255 & 10.73 \\
            4 & \verb+i32.const+ & 11,478,749 & 7.08 \\
            5 & \verb+loop,local.set+ & 9,959,227 & 6.14 \\
            6 & \verb+i32.load+ & 5,695,009 & 3.51 \\
            7 & \verb+loop,i32.const+ & 5,644,941 & 3.48 \\
            8 & \verb+i32.add+ & 3,524,522 & 2.17 \\
            9 & \verb+loop,i32.load+ & 3,514,771 & 2.16 \\
            10 & \verb+i32.store+ & 3,501,192 & 2.16 \\
            \hline
        \end{tabular}
    \end{minipage}%
    \begin{minipage}[t]{0.5\textwidth}
        \centering
        \caption{Distribution of least used Wasm instructions}
        \resizebox{\textwidth}{!}{%
        \begin{tabular}{ l l r r }
            \hline
            Wasm Instruction & Project Names & \# Files & \# Methods\\
            \hline
            \texttt{i32.atomic.store8} & liblnk-dev & 1 & 1 \\
            \texttt{f32.copysign} & libmeschach-dev & 1 & 1 \\
            \texttt{i32.atomic.rmw.xchg} & liblnk-dev & 1 & 1 \\
            \texttt{i64.rotr} & datamash & 1 & 1 \\
            \texttt{i64.extend32\_s} & libscca-dev & 1 & 1 \\
            \texttt{i32.atomic.rmw.sub} & liblnk-dev & 2 & 2 \\
            \texttt{i32.atomic.load} & liblnk-dev & 3 & 3 \\
            \texttt{i64.extend16\_s} & gir1.2-harfbuzz-0.0 & 3 & 10 \\
            \texttt{f32.nearest} & frei0r-plugins, libyaz5 & 4 & 4 \\
            \texttt{i32.rotr} & datamash, emboss, liblscp-dev, & 8 & 13 \\
            & liblzo2-2, libsane-common, libtiff-dev, \\
            & libyaml-cpp0.3-dev \\
            \texttt{i32.atomic.rmw.add} & gir1.2-harfbuzz-0.0, liblnk-dev & 10 & 28 \\
            \texttt{i64.trunc\_f32\_u} & freetds-bin, libamd2, libfst-dev, & 13 & 17 \\ 
            & liblouis-bin, libyaz5 \\
            \texttt{i32.atomic.load8\_u} & leveldb-doc, libfst-dev & 17 & 19 \\
            \texttt{i64.ctz} & gawk, libcapstone-dev, libfst-dev, & 18 & 21 \\
            & liblog4cplus-1.1-9 \\
            \hline
        \end{tabular}%
        }
        \label{table:least_common_paths}
    \end{minipage}
\end{table}

Three nonterminals can cause a nested structure in WAT ASTs: (1) \verb+block+, (2) \verb+if+-\verb+else+ conditionals, and (3) \verb+loop+. We define a \textit{nested path} as a sequence of these nonterminals and then a final terminal which reflects one of the Wasm instructions. Two examples of extracted raw nested paths are:
\begin{itemize}
    \item \verb+if,loop,block,if,f32.const+
    \item \verb+block,if,block,block,block,if,f64.ne+
\end{itemize}
We carry out two refinement steps to reduce the number of paths: 

\subsubsection{\label{sec:collapse}Collapsing repeating nonterminals}
When a nonterminal appears multiple times in a row in a sequence of nonterminals, we collapse it into a single occurrence. For instance, the second example given above will be transformed into \verb+block,if,block,if,f64.ne+, with the three adjacent \verb+block+ nonterminals combined into a single one. This process applies to nonterminals \verb+block+, \verb+if+, and \verb+loop+. This change reduces the number of paths to 40,654.

\subsubsection{\label{sec:drop}Dropping blocks}
One way of reducing paths even further is to drop all instances of one of the three nonterminals \verb+block+, \verb+if+, and \verb+loop+. Out of these three nonterminals, \verb+block+ carries the least semantic value, since \verb+if+ and \verb+loop+ are key to implementing conditional and iterative logic, respectively. We thus drop \verb+block+ nonterminals. After this refinement step, the total number of paths plummets to 3,352. This number is low enough for our purposes in devising a code representation, therefore we choose this number as the number of dimensions for our feature vectors.

As a separate study, we also dropped both \verb+if+ and \verb+loop+ subtrees (along with \verb+block+). That makes our paths analogous to the instructions themselves, without any nested structure. This aggressive dropping would lead to 185 paths, i.e. there would only be 185 instructions used in the training portion of our dataset.

Our refinement steps were primarily designed to make this work feasible by reducing the number of paths, rather than to enhance results. Evaluating their direct impact would require experiments using the 807,972 raw paths, which was not feasible due to the immense data volume: in our code embedding experiment (Section~\ref{sec:code-embedding}), the input vectors' size of 3,352 dimensions results in 4.84GB of data (and more for an in-memory representation). With raw paths, the data would span multiple TB, far beyond our infrastructure's RAM limitations.

Table~\ref{table:path_count} shows the most common paths. The most common path is \verb+local.get+, which appears when the program retrieves a variable's value. 
Table~\ref{table:least_common_paths} lists the least frequently encountered instructions, where the \verb+i32.atomic.*+ instruction family notably stands out as one of the least used. This rarity can be attributed to the specialized nature of these instructions that are designed for atomic operations. In fact, these specialized instructions are only available under the Wasm threads proposal\footnote{https://github.com/WebAssembly/threads/blob/main/proposals/threads/Overview.md}.

\begin{figure}[!t]
    \begin{minipage}[b]{0.5\textwidth}
        \hspace*{-1cm} 
        \includegraphics[scale=0.2]{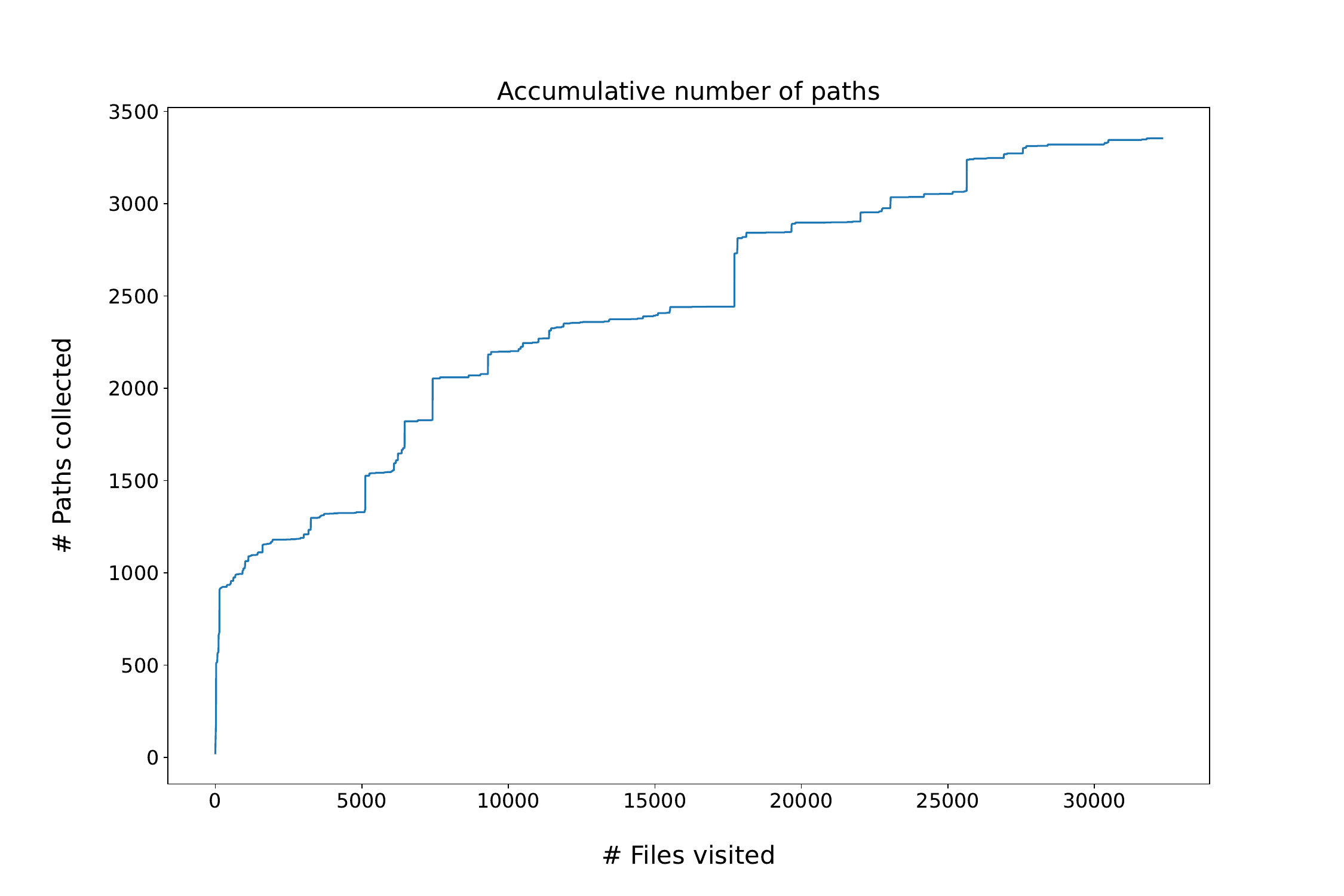}
        \captionof{figure}{Accumulative number of paths}
        \label{fig:accum}
    \end{minipage}%
    \begin{minipage}[b]{0.5\textwidth}
        \small
        \hspace*{-0.5cm} 
        \captionof{table}{Instruction present in WASI binaries but not \textsc{SnowWhite} binaries}
        \begin{tabular}{ l l }
            \hline
            \scriptsize{Wasm Instructions} & \scriptsize{Why missing in \textsc{SnowWhite}?} \\
            \hline
            \texttt{i16x8.*} & \scriptsize{Belongs to Wasm SIMD} \\
            \texttt{i32x4.*} & \scriptsize{proposal (not enabled in \textsc{SnowWhite})}\\
            \texttt{i64x2.*} & \\
            \texttt{i8x16.*} & \\
            \texttt{v128.*} & \\
            \hline 
            \texttt{i64.extend8\_s} & \scriptsize{Belongs to Sign-extension operations} \\
            \texttt{i64.extend32\_s} & \scriptsize{proposal (not enabled in \textsc{SnowWhite})}\\
            \hline
            \texttt{memory.grow} & \scriptsize{Exists in WASI \texttt{\$sbrk} method} \\
            \texttt{memory.size} & \scriptsize{which is not a user method}\\
            \hline \\
        \end{tabular}
        \label{table:missing_instructions}
    \end{minipage}
\end{figure}

Figure~\ref{fig:accum} shows the accumulative number of paths after visiting the files of our dataset, sorted by alphanumerical order of the filename. As the figure depicts, the number of paths quickly increases to 1000. Then, it keeps on going up until it reaches a total of 3,352 paths at the end. Also, there are sudden jumps in the accumulative plot, most notably around 17,500, suggesting that a small number of files can cause significant changes to our paths set. That indicates our data-driven path extraction approach requires a large dataset to produce reliable paths. To work with this constraint, we use the largest available WebAssembly dataset to obtain the most reliable paths set.

As discussed earlier, 3,352 is an acceptable number to set as the number of dimensions for our feature vectors. Each vector represents a function in a Wasm program and each element in the vector is an integer indicating the number of times the path associated with that element appears in the Wasm function's AST. 

\subsection{Verifying the paths with an alternative compilation toolchain}
\label{sec:alternative-compilation}
To confirm the generalizability of our Emscripten-based results, we selected 10 mid-to-large WASI-compliant Rust projects and compiled them to \verb+wasm32-wasi+ target. The obtained Wasm binaries contain over 9 million instructions and 42,356 methods. \verb+wasm32-wasi+ differs from an Emscripten target as it aims to provide a portable Wasm file that can be run by WebAssembly runtimes that support WASI standard (e.g. Wasmtime\footnote{https://github.com/bytecodealliance/wasmtime} or Wasmer\footnote{https://github.com/wasmerio/wasmer}). We intentionally used a different programming language (Rust) and a different compilation target (\verb+wasm32-wasi+) to demonstrate that our paths still apply in this broader setting.

Using WasmWalker, we collected paths from all methods within these 10 Wasm binaries, in contrast to the SnowWhite dataset, which samples methods from each binary file. We extracted 358 paths from a total of 42,356 methods. Out of these 358 paths, 67 (summarized in Table~\ref{table:missing_instructions}) were not included in the 3,352 paths set we collected by analyzing Emscripten-based binaries (and would be ignored by WasmWalker). However, out of these 67 paths, 65 did not appear in the SnowWhite dataset because the WASI binaries were compiled using the SIMD\footnote{https://github.com/zeux/wasm-simd/blob/master/Instructions.md} and Sign-extension operations\footnote{https://github.com/WebAssembly/threads/blob/main/proposals/sign-extension-ops/Overview.md} proposals, enabling new instructions (e.g., \verb+v128.load+, or \verb+i64.extend8_s+) in Wasm; the SnowWhite authors compiled their corpus without enabling these two proposals, so their compiled corpus contains no SIMD or Sign-extension instructions. The remaining two missing paths, \verb+memory.grow+ and \verb+memory.size+, exclusively appear in the \verb+$sbrk+ method, which is an embedded method within WASI binaries responsible for memory allocation. The SnowWhite dataset solely encompasses Wasm methods corresponding to user methods in C/C++, omitting the system-level methods embedded within the binaries by Emscripten, which use these additional two instructions. Thus, our analysis shows that every path in our user methods has been previously encountered in Emscripten-based Wasm binaries, as long as it does not use specialized instructions.

\subsection{Code representation}

Using our path vectors, we devise two different code representations. Our goal for devising these code representations is to use them as inputs to deep learning models---Section~\ref{sec:evaluation} discusses applications extensively. In the following sections, we describe each of these representations in detail.

\subsubsection{Path Sequence}
We can use a simple function $s$ to transform a path vector $v$ to a sequence $s(v)$ which contains all the necessary AST information. We define the function $s$ as follows:
\begin{equation*}
s: \mathbb{N}^{{3352}}\mapsto \{\langle n_1, m_1 \rangle, \ldots\}, 1 \leq n_1 \leq 3352, 1 \leq m_1 \leq D,
\end{equation*}
\begin{equation*}
s(v) :=  \{\langle e.\textit{index}, \lceil\frac{eD}{\sum\limits_{v_i\in v}{v_i}}\rceil \rangle\,\,|\,\,e \in v, e\neq0\}.
\end{equation*}

As shown in the equation above, each non-zero element in $v$ maps to a tuple $\langle n, m \rangle$. The first argument $n$ is the path's index in our indexed paths set, and the second argument $m$ shows the number of times the $n^{\text{th}}$ path appeared in the AST. In tuple $\langle n, m \rangle$, a high value for $m$ indicates that the $n^{\text{th}}$ path appeared many times in the AST. Note that we normalize the values of $m$ between 1 and $D$. We empirically found that setting $D=30$ is enough for our purposes. A high value for $D$ can create redundant symbols that could have formed a single symbol. To give an example, a vector like $v=[1,0,204,...,2]$ can be transformed to $s(v)=\langle 1,1 \rangle\,\,\langle 3,30 \rangle\,\,...\,\,\langle 3352,1 \rangle$. Note that the average number of unique paths that exist in a Wasm function is 21.5. That is, a path sequence has on average around 21 tuples.

\subsubsection{Code Embedding}
\label{sec:code-embedding}

An embedding is a mapping from objects to vectors of real numbers (represented on machines by floating-point numbers). Embeddings make it possible to work with textual data in a mathematical model. They also have the advantage that fundamentally discrete data (words) is transformed into a continuous vector space. In natural languages, embeddings can be created at different granularity levels, such as words, sentences, or documents. Similarly, in programming languages, we can create embeddings for program tokens, statements, or functions~\cite{chen2019literature}.

One of the goals of embedding is to make similar objects have similar vectors. For instance, in the Skip-gram model, the words that are used in similar contexts will have similar embeddings. This similarity, however, must be defined based on our objective. For example, two functions can be similar in ``method name'' but not similar in ``contain security vulnerability'' (which presumably depends on certain features of how the functions were implemented).

Neural networks are often used for creating embeddings because of scalability and the fact that they are capable of learning relationships between large numbers of words, making them well-suited for processing large-scale text datasets. Also, neural networks can model complex, non-linear relationships between words, allowing them to capture subtle relationships and associations.

Similar to word2vec and code2vec, we use simple feedforward neural networks to create code embeddings. However, unlike word2vec, we do not use the one-hot encoding~\cite{harris2012digital}, because one-hot encoding provides no measure of similarity between similar objects (e.g. words like ``apple'' and ``orange''). Our inital vectors encode AST paths, so similar objects might share similar paths. We use a similar network to the one used in code2vec to create our code embeddings. The only difference is that our method does not require having an attention weights vector, as each element in our input vector represents exactly one path. In code2vec, by contrast, each leaf-to-leaf path is encoded with a few hundred entries of the input vector. The attention weights vector is a vector in code2vec's neural architecture that contains a weight for each leaf-to-leaf path that corresponds to the path's impact in determining the output label.

%% file: evaluation.tex
While our primary objectives in this work are to contribute the notion of path-based representations for Wasm and that of extracting both fixed-sized and variable-sized representations from Wasm code, we do provide an comparative evaluation of our approach in cases where it can compare directly to previous work. We thus evaluate our approach on two tasks: (1) prediction of method names and (2) recovery of precise return types, using the same dataset for both evaluations. First, we provide an overview of the general characteristics of the dataset. Then we introduce the models that we created and used for evaluation. Finally, we present our results for method name prediction in Section~\ref{sec:method_name_prediction} and for precise return type recovery in Section~\ref{sec:recovering_type}.

\textbf{Dataset} We used the same dataset introduced in \textsc{SnowWhite}~\cite{lehmann2022finding} for three main reasons: (1) the dataset has been filtered and each data point is linked to a sequence of instructions, which are refined for more accurate type recovery. This provides us with a good opportunity to examine whether our insights regarding common AST paths and our path sequence can enhance accuracy results where both our approach and \textsc{SnowWhite} apply. (2) The dataset contains ground-truth DWARF debugging information, including function names and precise types, properties that our experiments are designed to predict. (3) To the best of our knowledge, this is the largest dataset for WebAssembly and is significantly larger than the datasets used in previous works~\cite{haas2017bringing, hilbig2021empirical, lehmann2020everything, jangda2019not}.

The dataset used in this study contains 6.3 million samples of WebAssembly code and type information obtained from 300,905 object files that were compiled from 4,081 C and C++ packages for the Ubuntu operating system. To prevent artificial inflation of results, the \textsc{SnowWhite} dataset curators applied some pre-processing steps, such as deduplication and discarding data points where the number of return types in WebAssembly does not match the number in C/C++ code. In addition to that, to use the wasm2wat module as part of deriving ASTs, we also discarded datapoints that take more than 30 seconds to translate to WAT files. The dataset has already been divided, based on the number of packages, into three portions, with 96\% for training, 2\% for early stopping and evaluating hyperparameters, and 2\% reserved for a held-out test set.

\textbf{Models} We trained five different models to assess whether AST paths enable us to create better models for predicting properties of Wasm programs.
To carry out a more accurate comparison with \textsc{SnowWhite}, we use the same sequence-to-sequence model as \textsc{SnowWhite} for predicting the types of function parameters and return values in code. The only difference between our approach and \textsc{SnowWhite} is that we did not use subword tokenization because it had zero impact on our model's performance. \textsc{SnowWhite} uses a bidirectional LSTM model~\cite{schuster1997bidirectional} with global attention~\cite{bahdanau2015neural} and dropout~\cite{srivastava2014dropout} for regularization. The model is optimized with backpropagation-through-time gradient descent using the Adam optimizer~\cite{kingma2015adam}. The \textsc{SnowWhite} authors chose hyperparameters including hidden vector dimension, number of layers, learning rate, dropout rate, and embedding dimension through experimentation. To conduct an accurate comparison, we use the same hyperparameters. The model has a total of 5.5 million learnable parameters. Both we and the \textsc{SnowWhite} authors also experimented with the Transformer architecture but found that the LSTM model was more effective. There are more computationally expensive LLM architectures, such as CodeBERT~\cite{feng-etal-2020-codebert}, that we could experiment with. However, this work focuses on the advantages of integrating path-awareness to Wasm code representations, and improving results in more resource-constrained environments. 

We use the same sequence-to-sequence architecture for our five models, with each model receiving a different input. The first model (seq2seq-INP) receives a concatenation of the last 20 Instructions and Nested Path sequences. (Capitalized letters indicate the abbreviation mapping, e.g. {\bf INP} stands for {\bf I}nstructions and {\bf N}ested {\bf P}aths.) Remember that we are concatenating two strings of different types: the sequence of the last 20 instructions (in textual, i.e. WAT, format), and a sequence of tuples representing the frequency of each nested path in our program's AST (similar to path\_sequence shown in ~\ref{fig:codes:representations}). We will justify our choice of the last 20 instructions in Section~\ref{sec:discussion}. The second model (seq2seq-ISP) is similar to the first model, except the path sequence does not reflect the 3,352 nested paths but 185 Simpler Paths after dropping \verb+if+ and \verb+loop+ subtrees, as discussed in Section~\ref{sec:methodology}. The third model (seq2seq-I) only receives the last 20 Instructions sequences, which is the model used in \textsc{SnowWhite}. The fourth model (seq2seq-NP) only receives the path sequence, reflecting the 3,352 Nested Paths. The fifth model (seq2seq-SP) is similar to the fourth model, but it uses the 185 Simpler Paths instead of the nested ones. 

We chose these five models to control the type and amount of information being input to our models and to enable accurate comparisons between them. The seq2seq-INP model receives the most information, as it takes in both the instruction and path sequences, with the path sequence being nested rather than simple.

Like \textsc{SnowWhite}, we used OpenNMT~\cite{klein2017opennmt} to train our models. We trained our models using Google Colaboratory Pro+. Our training setup was an Intel(R) Xeon(R) CPU with 3.7GHz clock speed and 16 cores, 85GB of RAM, and an NVIDIA P100 GPU with dedicated memory. Training of each model, including the \textsc{SnowWhite}-like model, took about the same amount of time, which was less than 10 minutes.

\subsection{\label{sec:method_name_prediction}Method name prediction}
Method name prediction provides a label that describes a method's semantics by analyzing its body. It can be especially useful for identifying malicious methods and reverse engineering. We now discuss how we evaluate the effectiveness of our code representation in method name prediction. First, to avoid manual inspection after model prediction, we carry out a preprocessing step to simplify method names. This preprocessing step includes: (1) removing generic types that a method name may include, (2) removing initial underscores that usually indicate private/protected methods, (3) converting name letters to lowercase. The preprocessing step also prevents artificial inflation of our accuracy scores and enables us to conduct a more meaningful analysis of our results. Next, we create different datasets parameterized by $m$, the minimum number of datapoints associated with a method name for it to be included in the dataset. For instance, if a dataset is created with $m=50$, that means each method name in the dataset has at least 50 associated datapoints. We create four datasets for $m=10 / 20 / 50 / 100$. The datasets include 345,417 / 279,262 / 223,903 / 192,905 datapoints and 7,560 / 2,768 / 874 / 412 unique method names, respectively.

\begin{table*}[]
    \centering
    \caption{Method name prediction accuracy results: The seq2seq-INP model provides the highest accuracy scores overall. Additionally, seq2seq-INP and seq2seq-ISP models outperform seq2seq-I, which was used in \textsc{SnowWhite}. Models without instruction sequences, seq2seq-NP and seq2seq-SP, generally yield lower accuracy compared to seq2seq-I.}
    \label{table:method_name} 
    \resizebox{\textwidth}{!}{
    \begin{tabular}{lrrrrrrrr}
    \multicolumn{1}{l}{} & \multicolumn{8}{c}{} \\
       \hline
     & \multicolumn{2}{l}{m=10} & \multicolumn{2}{l}{m=20} & \multicolumn{2}{l}{m=50} & \multicolumn{2}{l}{m=100} \\
     \hline
     & \multicolumn{1}{l}{top-1 acc.} & \multicolumn{1}{l}{top-5 acc.} & \multicolumn{1}{l}{top-1 acc.} & \multicolumn{1}{l}{top-5 acc.} & \multicolumn{1}{l}{top-1 acc.} & \multicolumn{1}{l}{top-5 acc.} & \multicolumn{1}{l}{top-1 acc.} & \multicolumn{1}{l}{top-5 acc.} \\
     \hline
     \hline \\
    seq2seq-INP & \textbf{75.97\%} & \textbf{92.79\%} & \textbf{77.03\%} & \textbf{94.67\%} & \textbf{76.82\%} & \textbf{96.01\%} & \textbf{76.62\%} & \textbf{96.68\%}\\
    seq2seq-ISP & 74.80\% & 92.68\% & 74.71\% & 93.59\% & 75.95\% & 95.59\% & 75.76\% & 96.07\%\\
    seq2seq-I & 74.19\% & 90.72\% & 73.68\% & 91.41\% & 71.46\% & 84.70\% & 74.25\% & 92.48\%\\
    seq2seq-NP & 71.54\% & 89.25\% & 72.11\% & 91.51\% & 71.53\% & 79.53\% & 72.75\% & 92.65\%\\
    seq2seq-SP & 71.13\% & 88.75\% & 72.70\% & 86.61\% & 70.96\% & 79.64\% & 64.21\% & 86.61\%\\
    \end{tabular}
    }
\end{table*}

Table~\ref{table:method_name} presents the prediction accuracy results after training each model on each dataset. \textbf{As shown in the table, our seq2seq-INP model offers the best accuracy scores across the board.} Moreover, both seq2seq-INP and seq2seq-ISP models offer better accuracy than seq2seq-I, which was the model used in \textsc{SnowWhite}. Notably, the best improvement occurs at $m=50$, with seq2seq-INP surpassing seq2seq-I by 5.36\% in top-1 accuracy and 11.31\% in top-5 accuracy.

Moving on to the models that were trained without being given instruction sequences, seq2seq-NP and seq2seq-SP usually resulted in less accuracy than seq2seq-I. Under some circumstances, omitting the instruction order does not affect the model accuracy. It's worth emphasizing that method name prediction was not included in the evaluation process of the original \textsc{SnowWhite} study. Thus, to conduct a more meaningful comparison in this experiment, we adapted their pipeline to predict method names instead of program types.

\begin{figure}[!t]
    \centering
    \includegraphics[scale=0.35]{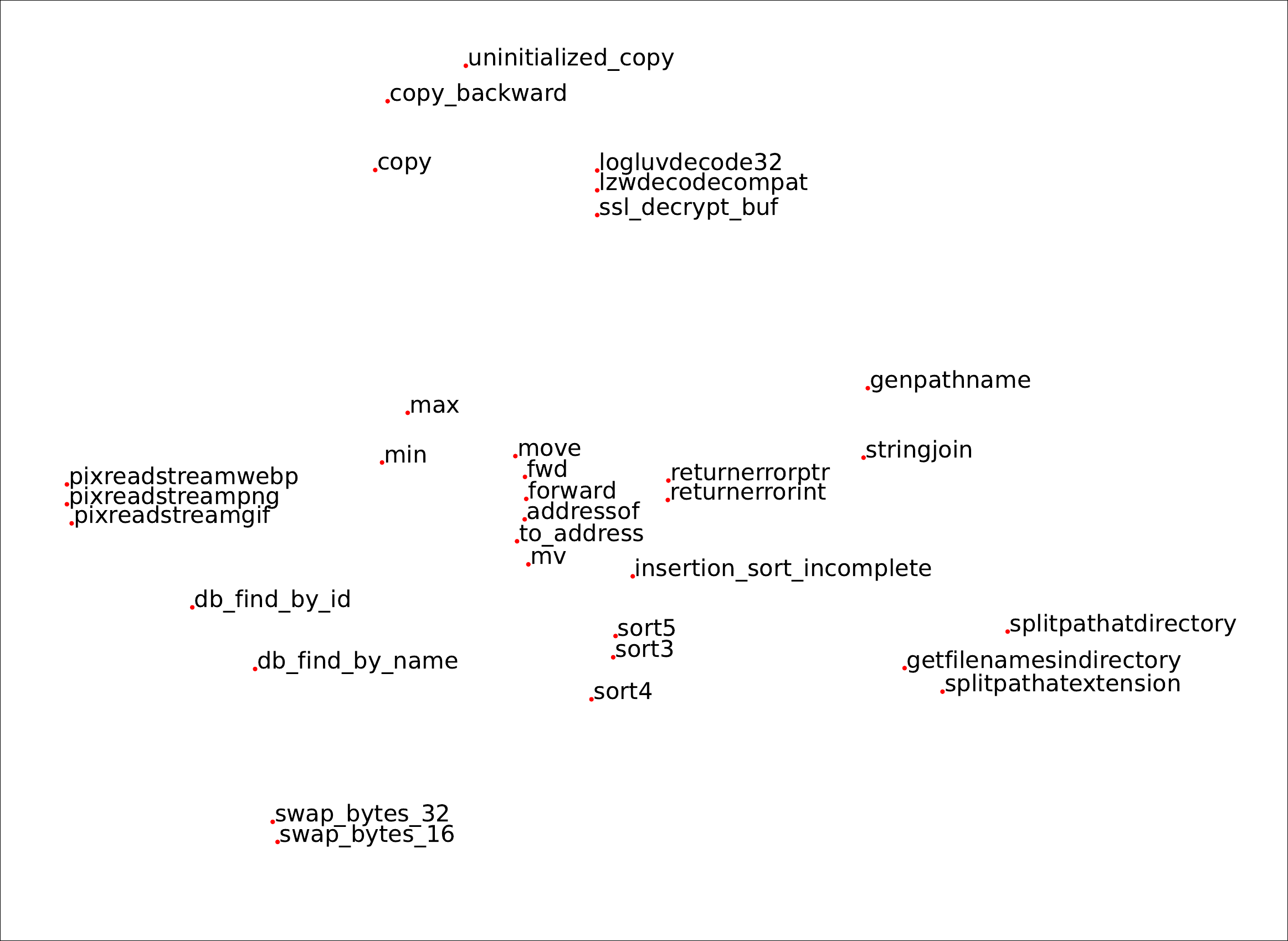}
    \caption{A t-SNE 2D plot of code embeddings generated using our proposed code embedding approach. The plot shows the spreading of method names in the 2D plane, with similar method names closer to each other, highlighting the effectiveness of our embedding approach in capturing the semantic similarities among Wasm program methods. To prevent label overlapping, we manually adjusted the vertical position of points by up to 3.1\%.}
    \label{fig:embedding}
\end{figure}

\subsubsection{Code embedding} As a side study, we created code embeddings for method names. Since our approach, unlike \textsc{SnowWhite}'s, is path-aware, we can leverage existing methods to generate code embeddings from our AST paths. Our goal in this study is to depict similar method names close to each other, providing evidence that the embeddings have the correct understanding of code semantics. Additionally, this would further validate that our path vectors appropriately embed AST information. Figure~\ref{fig:embedding} shows a 2D visualization of our embeddings using t-SNE~\cite{maaten2008visualizing}, a technique used for dimensionality reduction and visualization of high-dimensional datasets. The point associated with each method name shows the location of the average of all embeddings with that name. This visualization effectively conveys the relationships between method names and their semantics. Our path-based representation enables a user to inspect the similar paths in two methods that are flagged as being similar---they can understand some evidence as to why the methods were said to be similar.

To produce these embeddings, we used a feedforward network with the following characteristics: four hidden layers, which employ dropout and L2 kernel regularization to prevent overfitting; ReLU and softmax activation functions in the hidden and output layers, respectively; and the Adam optimizer. This model takes path vectors as input. The last hidden layer has 50 nodes that contain our embedding values. We chose $m = 10$.

With these embeddings, we are able to cross-check correlations between method names---names have independent semantic meanings. By examining these correlations, we aim to further validate our model's effectiveness in capturing underlying semantics. This additional analysis shows the quality of the embeddings we generate. We discuss some method name similarities in the figure:

\textit{A Closer Look:} ``lzwdecodecompat'' and ``logluvdecode32'' both decode compressed data. ``lzwdecodecompat'' decodes LZW (Lempel-Ziv-Welch) data compression, which is a lossless compression algorithm commonly used in the past for images, text, and other types of data. ``logluvdecode32'', on the other hand, decodes LogLuv-encoded image data. LogLuv is a non-linear color space that was developed for representing high-dynamic-range (HDR) image data, and it is used in many professional graphics applications. While these methods are used for different types of data compression, they are similar in that they are both used for decoding compressed data.

The figure also shows ``ssl\_decrypt\_buf'' close to ``lzwdecodecompat'' and ``logluvdecode32''. ``ssl\_decrypt\_buf'' is selected from a package related to SSL/TLS protocol, a cryptographic protocol that is used to provide secure communication over the internet. As the name suggests, ``ssl\_decrypt\_buf'' refers to decrypting SSL traffic, which is similar to ``lzwdecodecompat'' and ``logluvdecode32'', as all of these methods aim to retrieve the original encoding of a data stream that has been transformed to a different encoding.

More instances of method name pairs are 
``db\_find\_by\_name''/``db\_find\_by\_id'', ``min''/``max'', and ``swap\_bytes\_16''/``swap\_bytes\_32''.

\subsection{\label{sec:recovering_type}Recovering precise return types}

As we described earlier, Lehmann et al.~\cite{lehmann2022finding} proposed \textsc{SnowWhite}, the first tool for recovering precise types from Wasm binaries. The WebAssembly type system only includes four primitive types: ``i32'', ``i64'', ``f32'', and ``f64''. \textsc{SnowWhite} successfully showed that it is possible to recover precise high-level C/C++ types from Wasm primitive types using a seq2seq model that gets a portion of Wasm instructions as input. Their work used two different models to recover precise types for (1) method parameters, and (2) return values. Our initial assumption was that providing the model with AST knowledge would improve accuracy. Since the primary objective of this paper is not precise type recovery, but rather offering a novel code representation, we only evaluate recovering precise return types and not parameter types.

For evaluating \textsc{SnowWhite}, the authors defined a high-level type language that governed the to-be-predicted type sequences. Additionally, they defined five different type variants and, based on those variants, carried out five different evaluations: 
\begin{enumerate}
    \item  $\mathcal{L}_{\text{SW}}$: The default version that applies filtering on types (e.g. omitting static arguments of instructions that are likely unhelpful and unnecessarily). For instance, a type under this variant might look like ``\verb+pointer const primitive uint32_t+''.
    \item $\mathcal{L}_{\text{SW, All Names}}$: Disables all filters. An instance would be ``\verb+pointer name "SARRAY" struct+'', containing the actual type's name.
    \item $\mathcal{L}_{\text{SW, Simplified}}$: Removes names, constness and the distinction between classes and structs from $\mathcal{L}_{\text{SW}}$. The example in (1) would be ``\verb+pointer primitive uint32_t+'' under this variant.
    \item $\mathcal{L}_{\text{SW, $t_{\text{low}}$ not given}}$ : Each input sequence starts with the actual low-level WebAssembly return type (like \verb+i32+ or \verb+f64+). This variant is similar to $\mathcal{L}_{\text{SW}}$, except the low-level WebAssembly type is removed from the beginning of the input sequence.
    \item $\mathcal{L}_{\text{Eklavya}}$ : This variant is based on a prior work~\cite{chua2017neural} that has only seven primitive types (\verb+int+, \verb+float+, \verb+char+, \verb+pointer+, \verb+enum+, \verb+union+, \verb+struct+). The example in (1) would be simply ``\verb+pointer+'' under this variant.
\end{enumerate}
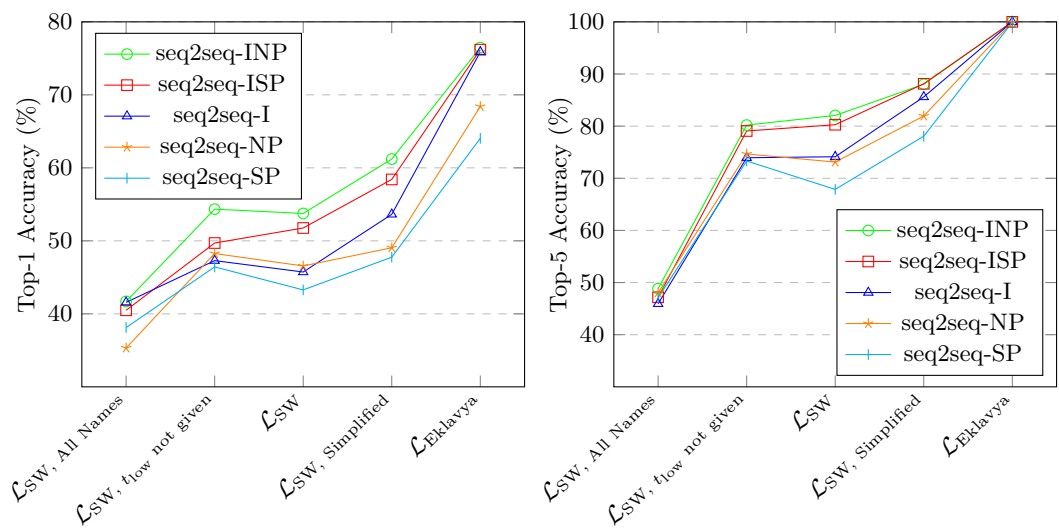
\begin{figure*}[h]    
    \begin{tikzpicture}
    \begin{axis}[
        scale=0.85,
        ylabel={Top-1 Accuracy (\%)},
        y label style={
            yshift=-15pt, 
        },
        font=\small,
        xmin=0.5, xmax=5.5,
        ymin=30, ymax=80,
        xtick={1,2,3,4,5},
        xticklabels={$\mathcal{L}_{\text{SW, All Names}}$,$\mathcal{L}_{\text{SW, $t_{\text{low}}$ not given}}$,$\mathcal{L}_{\text{SW}}$,$\mathcal{L}_{\text{SW, Simplified}}$,$\mathcal{L}_{\text{Eklavya}}$},
        x tick label style={
            rotate=45,
            anchor=east,
        },
        ytick={40,50,60,70,80,90},
        legend pos=north west,
        ymajorgrids=true,
        grid style=dashed,
    ]

    \addplot[
        color=green,
        mark=o,
        ]
        coordinates { 
        (1,41.70)
        (2,54.34)
        (3,53.75)
        (4,61.20)
        (5,76.47)
        };
        \addlegendentry{seq2seq-INP}
    
    \addplot[
        color=red,
        mark=square,
        ]
        coordinates { 
        (1,40.49)
        (2,49.70)
        (3,51.76)
        (4,58.41)
        (5,76.19)
        };
        \addlegendentry{seq2seq-ISP}
    
    \addplot[
        color=blue,
        mark=triangle,
        ]
        coordinates { 
        (1,41.57)
        (2,47.29)
        (3,45.73)
        (4,53.63)
        (5,75.88)
        };
        \addlegendentry{seq2seq-I}

    \addplot[
        color=orange,
        mark=star,
        ]
        coordinates { 
        (1,35.32)
        (2,48.26)
        (3,46.56)
        (4,49.05)
        (5,68.41)
        };
        \addlegendentry{seq2seq-NP}

    \addplot[
        color=cyan,
        mark=|,
        ]
        coordinates { 
        (1,38.12)
        (2,46.43)
        (3,43.28)
        (4,47.76)
        (5,64.04)
        };
        \addlegendentry{seq2seq-SP}

    \end{axis}


    \begin{axis}[
        scale=0.85,
        at={(0.5\textwidth,0)}, 
        ylabel={Top-5 Accuracy (\%)},
        y label style={
            yshift=-15pt, 
        },
        font=\small,
        xmin=0.5, xmax=5.5,
        ymin=30, ymax=100,
        xtick={1,2,3,4,5},
        xticklabels={$\mathcal{L}_{\text{SW, All Names}}$,$\mathcal{L}_{\text{SW, $t_{\text{low}}$ not given}}$,$\mathcal{L}_{\text{SW}}$,$\mathcal{L}_{\text{SW, Simplified}}$,$\mathcal{L}_{\text{Eklavya}}$},
        x tick label style={
            rotate=45,
            anchor=east,
        },
        ytick={40,50,60,70,80,90,100},
        legend pos=south east,
        ymajorgrids=true,
        grid style=dashed,
    ]
    \addplot[
        color=green,
        mark=o,
        ]
        coordinates { 
        (1,48.83)
        (2,80.20)
        (3,82.05)
        (4,88.03)
        (5,100.00)
        };
        \addlegendentry{seq2seq-INP}
    
    \addplot[
        color=red,
        mark=square,
        ]
        coordinates { 
        (1,47.19)
        (2,79.07)
        (3,80.27)
        (4,88.12)
        (5,100.00)
        };
        \addlegendentry{seq2seq-ISP}
    
    \addplot[
        color=blue,
        mark=triangle,
        ]
        coordinates { 
        (1,45.93)
        (2,73.94)
        (3,74.13)
        (4,85.60)
        (5,100.00)
        };
        \addlegendentry{seq2seq-I}

    \addplot[
        color=orange,
        mark=star,
        ]
        coordinates { 
        (1,48.03)
        (2,74.66)
        (3,73.11)
        (4,81.94)
        (5,99.95)
        };
        \addlegendentry{seq2seq-NP}

    \addplot[
        color=cyan,
        mark=|,
        ]
        coordinates { 
        (1,47.08)
        (2,73.32)
        (3,67.87)
        (4,78.06)
        (5,99.91)
        };
        \addlegendentry{seq2seq-SP}
    \end{axis}

    \end{tikzpicture}   
    \caption{\label{fig:type} Model Accuracies for precise return type recovery}
    
\end{figure*}

\newpage 
Figure~\ref{fig:type} compares the accuracy scores of our five models. Recall that seq2seq-I is our reproduction of \textsc{SnowWhite} while seq2seq-INP is our most sophisticated model, incorporating both the last-20 instructions sequence and the nested path sequences information. \textbf{Like we found with method name prediction, our concatenation of instructions and path sequences (seq2seq-INP) resulted in the highest accuracy scores both in Top-1 and Top-5 accuracy plots.} More specifically, seq2seq-INP resulted in better accuracy than seq2seq-I by 8.02\% top-1 accuracy and 7.92\% top-5 accuracy in $\mathcal{L}_{\text{SW}}$. Similarly, seq2seq-ISP yields better accuracy than seq2seq-I by 6.02\% top-1 accuracy and 6.14\% top-5 accuracy in $\mathcal{L}_{\text{SW}}$. This improvement in accuracy suggests that our approach can predict richer and more nuanced types. Cross-checking our reproduction of \textsc{SnowWhite}, the accuracy scores displayed for seq2seq-I are close to those reported by the \textsc{SnowWhite}. Minor variations may exist due to missing training data points caused by the 30-second timeout during WAT generation.

As with method name prediction, seq2seq-NP, a model that was not given the instruction sequences as input, offers better accuracy than seq2seq-I in top-1 accuracy in $\mathcal{L}_{\text{SW, $t_{\text{low}}$ not given}}$ and $\mathcal{L}_{\text{SW}}$, and in top-5 accuracy in $\mathcal{L}_{\text{SW, All Names}}$ and $\mathcal{L}_{\text{SW, $t_{\text{low}}$ not given}}$.

For both method name prediction and precise return type recovery, nested paths models (seq2seq-INP and -NP) give better accuracy than their simple paths counterparts (seq2seq-ISP and -SP).

%% file: discussion.tex
We began by extracting empirical insights from the nested structures of WAT files. These findings not only served as the foundation and main motivation for WasmWalker's development, but also provide valuable insights in their own right. For instance, the list of least common Wasm instructions within compiled Ubuntu packages can directly assist Wasm runtime developers in assessing the compatibility of their runtimes with real-world binaries. WasmWalker is built upon empirical data comprising all AST paths found in the largest available Wasm corpus. This collected set of AST paths also serves as valuable empirical data, shedding light on the structure of WAT files. However, our goal was not merely to collect empirical data. Instead, we aimed to leverage this empirical information to create useful tools that contribute to the analysis of Wasm binaries.

Our hypothesis was that information about AST paths in Wasm can improve the effectiveness of our models. We tested this hypothesis by applying our models to two different tasks: (1) method name prediction, and (2) recovering precise return types. Our results demonstrate the adaptability and versatility of our AST-aware approach---there are fundamental differences in tool requirements for these two tasks---and also show its effectiveness. Additionally, we showed our technique's effectiveness on a third task: the AST paths enable effective code embedding, which is challenging for non-path-aware approaches like \textsc{SnowWhite}.

Method name prediction and return type recovery benefit from different types of information about the program under analysis. Method name prediction usually requires analysis of the program's control flow, while data flow analysis ought to be more helpful for recovering precise high-level types. All three tasks can be helpful for reverse engineering, detecting security vulnerabilities, and understanding the internal structures used inside the program.

Another key difference between tasks is that method name prediction and code embedding conceptually should require all of the (binary) code of the target function, i.e. a method name is associated with the whole method body. On the other hand, recovering precise return types can definitely be done with a selected portion of the binary code of the target function. In practice, recall that we found that method name prediction works to some extent with the last 20 instructions, but adding the paths (INP vs I) works 11.31\% better (top-5) on our dataset for method name prediction.

Despite the differences between the tasks of method name prediction and recovering precise high-level types, we showed that both tasks benefit from knowledge of the AST. More specifically, we observed that using the seq2seq-INP model, which includes nested AST paths information in the code representation, we achieve a higher accuracy compared to the \textsc{SnowWhite}-like seq2seq-I model, which does not. This improvement in accuracy was observed across a diverse set of programs in our dataset, containing Ubuntu packages with a wide range of software, including networking programs and high-performance security and cryptographic packages. \textbf{Therefore, incorporating AST paths information can be beneficial for both method name prediction and recovering precise high-level types, and this holds true for a wide variety of programs; AST paths also enable further applications like code embedding.}

Providing more instructions could also be a way to embed more knowledge about the program in our inputs. However, this would not be practical, given that even a basic Wasm program has thousands of instructions. Our novelty lies in our ability to store the AST information using a concise path sequence. 

WasmWalker provides ways to construct both fixed-sized and variable-sized inputs for a multitude of predictive models. On the fixed-sized side, unlike \textsc{SnowWhite}, our pipeline can produce code embeddings that can be used in traditional machine learning models like ANNs; and, the plot in Figure~\ref{fig:embedding} illustrates the effectiveness of our embeddings in grouping similar programs together.  \textsc{SnowWhite} is simply not usable for applications that require code embeddings. Regarding variable-sized inputs, our results offer compelling evidence that augmenting the sequence of instructions with our novel path sequence, which summarizes the structure of the AST in a compact form, leads to measurable improvements. 

A reader may wonder about the performance of the seq2seq-NP model without the I. Indeed, that model alone performs worse than seq2seq-I. We are proposing the joint use of the Instructions and the Nested Paths for variable-sized inputs; there is no good reason to use the Nested Paths alone for this type of input, and using both does better than just the Instructions. We reiterate that seq2seq-I, which is our replication of \textsc{SnowWhite}, is not path-aware and cannot be used in applications that work with code embeddings.

We incorporated concrete knowledge of Wasm binaries into the seq2seq-INP, -ISP and -I models by using the last 20 instructions sequence; we now discuss our two main reasons for choosing the specific number 20. Firstly, this was the approach used in previous work (\textsc{SnowWhite}) for recovering precise return types: the assumption was that the last 20 instructions are most closely related to the return type. Secondly, we needed to limit the number of instructions used in our models: using all instructions can result in overfitting and long training times. Giving more instructions to seq2seq models would lead to a linear increase in the additional space ($\Theta(n)$) proportional to the number of instructions $n$. On the other hand, the benefits of adding path sequences are independent of the selected number of instructions: our path sequence imposes a constant additional space complexity ($\Theta(1)$). That is also the reason why there is no noticeable difference in training time and model sizes compared to \textsc{SnowWhite}. We point out that the average number of unique paths in a path sequence is 21.5, and that the path sequence represents an entire function, not just the last 20 instructions.

We thus argue that incorporating structured AST knowledge is an economical and efficient method of enhancing our analysis models with more valuable information versus other linear approaches. While we designed WasmWalker for WebAssembly, the fundamental methodology is flexible and would apply to related technologies (like LLVM IR, or Java bytecode).


It is worth noting that our path extraction method did not include concrete terminals, as we aimed to reduce the number of paths and keep them simple. Additionally, the use of terminals can vary significantly for semantically similar programs. However, further investigation of how the terminals, mainly memory locations, change in Wasm binaries with slight variations in the high-level source code could improve our method. Such a study would provide more insight into how our model can leverage terminal information in its path extraction process.

\subsection{Threats to Validity} First, an overarching threat is that our dataset is not representative. The number of AST paths that we derived and used as the size of our feature vectors was based on an empirical study. Arbitrary Wasm programs may contain paths that are not included in our paths set. We mitigate this threat by using a large dataset. Given a new dataset, it is a straightforward task to recompute paths; while the number 3352 might change (especially in the presence of new instructions), we do not expect this to affect the viability of our representations. Second, our primary dataset consists of Wasm binaries that were compiled from C/C++ programs to emscripten target. We verified that changing the source code language and Wasm target did not introduce unforeseen paths by extracting paths from 10 Wasm binaries compiled from Rust projects using \verb+wasm32-wasi+ target, which differs from emscripten. While we have reason to believe that all Wasm compilers that use LLVM as part of their compilation pipeline (as do emscripten and \verb+wasm32-wasi+ compilation) would create the same paths, it is unclear if compilers that use a non-LLVM backend for creating binaries would yield the same set of paths. Finally, WasmWalker embodies a dependency on the current version of WebAssembly. As our supplementary study shows, if WebAssembly gains new features causing changes in AST strucures, we need to recollect paths and update our paths set accordingly---a routine non-research task.

%% file: related.tex
The analysis of binary code is a widely used method for studying closed-source programs. This approach has a wide range of applications, including identifying plagiarism, predicting performance, detecting malware, and discovering vulnerabilities. Researchers have developed many techniques for analyzing binary code.

Lehmann et al.~\cite{lehmann2022finding} proposed \textsc{SnowWhite}, a neural approach for recovering precise high-level parameter and return types of WebAssembly functions. To represent high-level types, \textsc{SnowWhite} uses an expressive type language that describes a large number of complex types, and builds on the success of neural sequence-to-sequence models for sequence prediction. \textsc{SnowWhite} is the first work that evaluates its performance on a large-scale dataset of 6.3 million type samples extracted from 300,905 WebAssembly object files. The results highlight that \textsc{SnowWhite}'s type language is expressive and accurately matches a significant portion of parameter and return types with the top-1 and top-5 predictions. Our work uses the dataset from \textsc{SnowWhite}. We also take advantage of the instruction sequences that our work provides to create a more effective code representation for WebAssembly functions which includes both the last 20 instructions as well as nested paths. A key difference between our approach and \textsc{SnowWhite}'s is that our representation is path-aware.

Feng et al.~\cite{feng2016scalable} proposed Genius, a graph embedding pipeline. Genius splits native code into a collection of basic blocks and creates an attributed control flow graph (ACFG); ACFG nodes are basic blocks. These nodes contain two types of attributes: block-level attributes (e.g. the number of instructions) and inter-block attributes (e.g. the number of offspring). To transform an ACFG to an embedding, Genius trains a codebook using a clustering algorithm. Then it leverages a bipartite graph matching algorithm to measure the similarity between an ACFG and the representative of a cluster. These similarity measures then map to fixed-sized feature vectors.

The main drawback of the Genius work is that generating the codebook is costly, and the quality of the codebook is limited by the size of the training dataset. Xu et. al~\cite{xu2017neural} improved on Genius with neural network-based embedding generation instead of bipartite graph matching. They tested their method on three different datasets, including the one from Genius, and achieved better accuracy and training time. Our work is similar to these two studies: we also transform block-level information into fixed-sized vectors. Like Xu et. al, we use a neural approach that yields a reduced training time compared to traditional methods like Genius codebooks. However, there are key differences. First, their ACFG graph traversal algorithm for generating embeddings does not involve path extraction, whereas our method is based on gathering paths from ASTs. Second, their dataset did not include Wasm binaries.

To capture code semantics, Ben-Nun et al.~\cite{ben2018neural} follow a hypothesis for computer programs akin to the linguistic Distributional Hypothesis~\cite{raychev2014code}. This hypothesis states that statements used in the same context often have similar semantics. They thus devise a representation for LLVM IR statements called conteXtual Flow Graphs (XFG) that incorporates the relative data- and control-flow of a statement. To build a XFG, they use LLVM IR statements in Static Single Assignment format. After preprocessing, they used the skip-gram model~\cite{mikolov2013distributed} to derive embeddings. The authors evaluated the derived XFG embeddings in three high-level classification and prediction tasks and found that they outperformed all manually extracted features. The results were comparable to or better than results from two inherently different specialized DNN solutions.

Similarly, Redmond et al.~\cite{redmond2018cross} introduced an instruction embedding model. They created a joint model combining the context information found in instruction sequences within the same architecture and the semantic similarities found in instruction sequence pairs from different architectures. They used word2vec's CBOW algorithm and evaluated the model on three tasks: (1) determining instruction similarity within the same architecture, (2) determining instruction similarity across different architectures, and (3) comparing basic block similarity across different architectures.

Our embeddings, unlike prior work, maps functions to vectors, rather than instructions to vectors. Furthermore, our results on Wasm ASTs provide not just an embedding, but also an interpretable sequence representation of a Wasm function, which is not offered by the prior studies. Our path sequence is interpretable as we draw each path from an indexed paths set (shown in Figure~\ref{fig:overview}).

Alon et al.~\cite{alon2019code2vec} proposed code2vec as a method to embed code in high-level programming languages into fixed-sized vectors. Their approach involved extracting paths from the AST, focusing specifically on leaf-to-leaf paths. They asserted that these paths contain more semantic information, as they include all the information between two terminals such as variable names in the AST. As discussed in Section~\ref{sec:methodology}, we believe that leaf-to-leaf paths do not contain additional useful semantic information in our context. They assigned fixed-sized vectors to paths and program tokens, which were then fed into a neural network to learn how to combine the vectors into a single embedding. The authors trained the network with an attention mechanism for predicting method names in Java programs and achieved promising results. However, code2vec faces challenges in creating input for the network. code2vec cannot handle the variable number of paths in a program, and additionally will always consider a quadratic number of leaf-to-leaf paths (which is why we were unable to directly compare our method with code2vec). These challenges can be addressed by adding padding or using RNNs to obtain a compressed representation of paths. In contrast, our method avoids these challenges, as it is based on the limited number of refined paths in WAT ASTs and the fact that we use root-to-leaf paths to avoid the quadratic explosion.

Like us, the ASTMiner~\cite{kovalenko2019pathminer} and PSIMiner~\cite{spirin_psiminer} projects also provide tools that extract representations from code for use in machine learning models. One of our primary contributions is the path-based refinements described in Section~\ref{sec:common-paths}. While PSIMiner's tree transformation framework appears to be able to support such refinements, they demonstrate simpler refinements like hiding literals and removing comments. Furthermore, both ASTMiner and PSIMiner work at AST level rather than at IR level as we do. Supporting WAT in these tools and evaluating our refinements in their context seems to require significant implementation effort. The benefits of integrating our tool with theirs also appear to be limited: like ASTMiner and PSIMiner, WasmWalker already generates output that is easily usable by the Python tools usually used as the next stage of machine learning pipelines.

%% file: conclusion.tex
In this study, using our pipeline, WasmWalker, we conducted the first empirical investigation of the common paths within the AST of Wasm programs over a large dataset of Ubuntu 18.04 packages. After preprocessing the collected paths, our analysis revealed that there are only 3,352 root-to-leaf paths within the ASTs of Wasm programs. Our main contribution lies in the development of two code representations based on our empirical findings about these 3,352 root-to-leaf paths that facilitate program analysis over WebAssembly binaries, providing researchers with a valuable tool for this purpose. The two code representations we propose are: (1) a path sequence that shows the paths inside a Wasm function and the number of times they occurred in the AST, and (2) a code embedding that encapsulates all the information within a Wasm function into a compact vector. To demonstrate the utility of our proposed code representations for WebAssembly program analysis, we evaluated them using five sequence-to-sequence models across two tasks: (1) method name prediction and (2) recovering precise return types. When concatenated with a portion of actual instructions, our path sequence representation led to improved prediction accuracy in both of these tasks. Specifically, it resulted in up to 5.36\% (11.31\%) improvement in Top-1 (Top-5) accuracy in method name prediction and 8.02\% (7.92\%) improvement in recovering precise return types, compared to the state-of-the-art. As a third task, our code embedding approach successfully provides similar embeddings for method names that have semantically similar method bodies. Although we tailored WasmWalker specifically for WebAssembly, the underlying approach can be readily adapted to other technologies.

%% file: main.bbl
\begin{thebibliography}{10}

\bibitem{emscripten}
Emscripten: {A} {C/C++} to {JavaScript} compiler.
\newblock \url{https://emscripten.org/}, 2023.
\newblock Accessed: March 3, 2023.

\bibitem{allamanis2015suggesting}
Miltiadis Allamanis, Earl~T Barr, Christian Bird, and Charles Sutton.
\newblock Suggesting accurate method and class names.
\newblock In {\em Proceedings of the 2015 10th joint meeting on Foundations of
  Software Engineering}, pages 38--49, 2015.

\bibitem{alon2019code2vec}
Uri Alon, Meital Zilberstein, Omer Levy, and Eran Yahav.
\newblock code2vec: {L}earning distributed representations of code.
\newblock {\em Proceedings of the ACM on Programming Languages}, 3(POPL):1--29,
  2019.

\bibitem{aumpansub2022learning}
Amy Aumpansub and Zhen Huang.
\newblock Learning-based vulnerability detection in binary code.
\newblock In {\em 2022 14th International Conference on Machine Learning and
  Computing (ICMLC)}, pages 266--271, 2022.

\bibitem{bahdanau2015neural}
Dzmitry Bahdanau, Kyunghyun Cho, and Yoshua Bengio.
\newblock Neural machine translation by jointly learning to align and
  translate.
\newblock In {\em Proceedings of the International Conference on Learning
  Representations (ICLR)}, 2015.

\bibitem{basque2024ahoy}
Zion~Leonahenahe Basque, Ati~Priya Bajaj, Wil Gibbs, Jude O’Kain, Derron
  Miao, Tiffany Bao, Adam Doup{\'e}, Yan Shoshitaishvili, and Ruoyu Wang.
\newblock Ahoy {SAILR}! there is no need to dream of {C}: A compiler-aware
  structuring algorithm for binary decompilation.
\newblock In {\em 33rd USENIX Security Symposium (USENIX Security 24)}, 2024.

\bibitem{batur2021novel}
Canan Batur~{\c{S}}ahin and Laith Abualigah.
\newblock A novel deep learning-based feature selection model for improving the
  static analysis of vulnerability detection.
\newblock {\em Neural Computing and Applications}, 33(20):14049--14067, 2021.

\bibitem{ben2018neural}
Tal Ben-Nun, Alice~Shoshana Jakobovits, and Torsten Hoefler.
\newblock Neural code comprehension: {A} learnable representation of code
  semantics.
\newblock {\em Advances in Neural Information Processing Systems}, 31, 2018.

\bibitem{brown2020language}
Tom~B Brown, Benjamin Mann, Nick Ryder, Melanie Subbiah, Jared Kaplan, Prafulla
  Dhariwal, Arvind Neelakantan, Pranav Shyam, Girish Sastry, Amanda Askell,
  et~al.
\newblock Language models are few-shot learners.
\newblock {\em arXiv preprint arXiv:2005.14165}, 2020.

\bibitem{buch2019learning}
Lutz B{\"u}ch and Artur Andrzejak.
\newblock Learning-based recursive aggregation of abstract syntax trees for
  code clone detection.
\newblock In {\em 2019 IEEE 26th International Conference on Software Analysis,
  Evolution and Reengineering (SANER)}, pages 95--104. IEEE, 2019.

\bibitem{chen2019literature}
Zimin Chen and Martin Monperrus.
\newblock A literature study of embeddings on source code.
\newblock {\em arXiv preprint arXiv:1904.03061}, 2019.

\bibitem{chua2017neural}
Zheng~Leong Chua, Shiqi Shen, Prateek Saxena, and Zhenkai Liang.
\newblock Neural nets can learn function type signatures from binaries.
\newblock In {\em USENIX Security Symposium}, pages 99--116, 2017.

\bibitem{cifuentes94:_rever}
Cristina Cifuentes.
\newblock {\em Reverse compilation techniques}.
\newblock PhD thesis, Queensland University of Technology, 1994.

\bibitem{devlin2017semantic}
Jacob Devlin, Jonathan Uesato, Rishabh Singh, and Pushmeet Kohli.
\newblock Semantic code repair using neuro-symbolic transformation networks.
\newblock {\em arXiv preprint arXiv:1710.11054}, 2017.

\bibitem{feng2016scalable}
Qian Feng, Rundong Zhou, Chengcheng Xu, Yao Cheng, Brian Testa, and Heng Yin.
\newblock Scalable graph-based bug search for firmware images.
\newblock In {\em Proceedings of the 2016 ACM SIGSAC Conference on Computer and
  Communications Security}, pages 480--491, 2016.

\bibitem{feng-etal-2020-codebert}
Zhangyin Feng, Daya Guo, Duyu Tang, Nan Duan, Xiaocheng Feng, Ming Gong, Linjun
  Shou, Bing Qin, Ting Liu, Daxin Jiang, and Ming Zhou.
\newblock {C}ode{BERT}: A pre-trained model for programming and natural
  languages.
\newblock In Trevor Cohn, Yulan He, and Yang Liu, editors, {\em Findings of the
  Association for Computational Linguistics: EMNLP 2020}, pages 1536--1547,
  Online, November 2020. Association for Computational Linguistics.
\newblock URL: \url{https://aclanthology.org/2020.findings-emnlp.139}, \href
  {https://doi.org/10.18653/v1/2020.findings-emnlp.139}
  {\path{doi:10.18653/v1/2020.findings-emnlp.139}}.

\bibitem{haas2017bringing}
Andreas Haas, Andreas Rossberg, Derek~L Schuff, Ben~L Titzer, Michael Holman,
  Dan Gohman, Luke Wagner, Alon Zakai, and JF~Bastien.
\newblock Bringing the web up to speed with {WebAssembly}.
\newblock In {\em Proceedings of the 38th ACM SIGPLAN Conference on Programming
  Language Design and Implementation}, pages 185--200, 2017.

\bibitem{harris2012digital}
David~Money Harris and Sarah~L. Harris.
\newblock {\em Digital Design and Computer Architecture}.
\newblock Morgan Kaufmann, 2012.

\bibitem{hilbig2021empirical}
Aaron Hilbig, Daniel Lehmann, and Michael Pradel.
\newblock An empirical study of real-world {WebAssembly} binaries: {S}ecurity,
  languages, use cases.
\newblock In {\em Proceedings of the Web Conference 2021}, pages 2696--2708,
  2021.

\bibitem{hochreiter1997long}
Sepp Hochreiter and J{\"u}rgen Schmidhuber.
\newblock Long short-term memory.
\newblock {\em Neural computation}, 9(8):1735--1780, 1997.

\bibitem{llvm}
The {LLVM}~Compiler Infrastructure.
\newblock {LLVM}: a compilation framework.
\newblock \url{https://llvm.org/}, 2022.
\newblock [Online; accessed 3 March 2023].

\bibitem{jangda2019not}
Abhinav Jangda, Bobby Powers, Emery~D Berger, and Arjun Guha.
\newblock Not so fast: {A}nalyzing the performance of {WebAssembly} vs. native
  code.
\newblock In {\em USENIX Annual Technical Conference}, pages 107--120, 2019.

\bibitem{kingma2015adam}
Diederik~P Kingma and Jimmy Ba.
\newblock Adam: {A} method for stochastic optimization.
\newblock In {\em Proceedings of the International Conference on Learning
  Representations (ICLR)}, 2015.

\bibitem{klein2017opennmt}
Guillaume Klein, Yoon Kim, Yuntian Deng, Jean Senellart, and Alexander~M Rush.
\newblock Opennmt: {O}pen-source toolkit for neural machine translation.
\newblock In {\em Proceedings of the Annual Meeting of the Association for
  Computational Linguistics: System Demonstrations}, pages 67--72, 2017.

\bibitem{kovalenko2019pathminer}
Vladimir Kovalenko, Egor Bogomolov, Timofey Bryksin, and Alberto Bacchelli.
\newblock Pathminer: a library for mining of path-based representations of
  code.
\newblock In {\em Proceedings of the 16th International Conference on Mining
  Software Repositories}, pages 13--17. IEEE Press, 2019.

\bibitem{lehmann2020everything}
Daniel Lehmann, Johannes Kinder, and Michael Pradel.
\newblock Everything old is new again: {B}inary security of {WebAssembly}.
\newblock In {\em Proceedings of the 29th USENIX Conference on Security
  Symposium}, pages 217--234, 2020.

\bibitem{lehmann2022finding}
Daniel Lehmann and Michael Pradel.
\newblock Finding the {D}warf: {R}ecovering precise types from {WebAssembly}
  binaries.
\newblock In {\em Proceedings of the 43rd ACM SIGPLAN International Conference
  on Programming Language Design and Implementation}, pages 410--425, 2022.

\bibitem{li2017cclearner}
Liuqing Li, He~Feng, Wenjie Zhuang, Na~Meng, and Barbara Ryder.
\newblock {CCLearner}: {A} deep learning-based clone detection approach.
\newblock In {\em 2017 IEEE International Conference on Software Maintenance
  and Evolution (ICSME)}, pages 249--260. IEEE, 2017.

\bibitem{li2018vuldeepecker}
Zhen Li, Deqing Zou, Shouhuai Xu, Xinyu Ou, Hai Jin, Sujuan Wang, Zhijun Deng,
  and Yuyi Zhong.
\newblock Vuldeepecker: {A} deep learning-based system for vulnerability
  detection.
\newblock {\em arXiv preprint arXiv:1801.01681}, 2018.

\bibitem{maaten2008visualizing}
Laurens van~der Maaten and Geoffrey Hinton.
\newblock Visualizing data using {t-SNE}.
\newblock {\em Journal of machine learning research}, 9(Nov):2579--2605, 2008.

\bibitem{manjula2019deep}
C~Manjula and Lilly Florence.
\newblock Deep neural network based hybrid approach for software defect
  prediction using software metrics.
\newblock {\em Cluster Computing}, 22(Suppl 4):9847--9863, 2019.

\bibitem{mikolov2013distributed}
Tomas Mikolov, Ilya Sutskever, Kai Chen, Greg~S Corrado, and Jeff Dean.
\newblock Distributed representations of words and phrases and their
  compositionality.
\newblock {\em Advances in neural information processing systems}, 26, 2013.

\bibitem{binaryen}
Mozilla.
\newblock Binaryen.
\newblock \url{https://github.com/WebAssembly/binaryen}, 2021.
\newblock Accessed: March 3, 2023.

\bibitem{qiao2020deep}
Lei Qiao, Xuesong Li, Qasim Umer, and Ping Guo.
\newblock Deep learning based software defect prediction.
\newblock {\em Neurocomputing}, 385:100--110, 2020.

\bibitem{raychev2014code}
Veselin Raychev, Martin Vechev, and Eran Yahav.
\newblock Code completion with statistical language models.
\newblock In {\em Proceedings of the 35th ACM SIGPLAN conference on Programming
  Language Design and Implementation}, pages 419--428, 2014.

\bibitem{redmond2018cross}
Kimberly Redmond, Lannan Luo, and Qiang Zeng.
\newblock A cross-architecture instruction embedding model for natural language
  processing-inspired binary code analysis.
\newblock {\em arXiv preprint arXiv:1812.09652}, 2018.

\bibitem{schuster1997bidirectional}
Mike Schuster and Kuldip~K Paliwal.
\newblock Bidirectional recurrent neural networks.
\newblock {\em IEEE Transactions on Signal Processing}, 45(11):2673--2681,
  1997.

\bibitem{spirin_psiminer}
Egor Spirin, Egor Bogomolov, Vladimir Kovalenko, and Timofey Bryksin.
\newblock Psiminer: A tool for mining rich abstract syntax trees from code.
\newblock In {\em 2021 IEEE/ACM 18th International Conference on Mining
  Software Repositories (MSR)}, pages 13--17, 2021.
\newblock \href {https://doi.org/10.1109/MSR52588.2021.00014}
  {\path{doi:10.1109/MSR52588.2021.00014}}.

\bibitem{srivastava2014dropout}
Nitish Srivastava, Geoffrey Hinton, Alex Krizhevsky, Ilya Sutskever, and Ruslan
  Salakhutdinov.
\newblock Dropout: {A} simple way to prevent neural networks from overfitting.
\newblock {\em Journal of Machine Learning Research}, 15(1):1929--1958, 2014.

\bibitem{wabt}
{The Wasm Team}.
\newblock {WebAssembly} {Binary} {Toolkit}.
\newblock \url{https://github.com/WebAssembly/wabt}, 2021.

\bibitem{vaswani2017attention}
Ashish Vaswani, Noam Shazeer, Niki Parmar, Jakob Uszkoreit, Llion Jones,
  Aidan~N Gomez, {\L}ukasz Kaiser, and Illia Polosukhin.
\newblock Attention is all you need.
\newblock In {\em Advances in Neural Information Processing Systems}, pages
  5998--6008, 2017.

\bibitem{wang2020detecting}
Wenhan Wang, Ge~Li, Bo~Ma, Xin Xia, and Zhi Jin.
\newblock Detecting code clones with graph neural network and flow-augmented
  abstract syntax tree.
\newblock In {\em 2020 IEEE 27th International Conference on Software Analysis,
  Evolution and Reengineering (SANER)}, pages 261--271. IEEE, 2020.

\bibitem{xu2017neural}
Xiaojun Xu, Chang Liu, Qian Feng, Heng Yin, Le~Song, and Dawn Song.
\newblock Neural network-based graph embedding for cross-platform binary code
  similarity detection.
\newblock In {\em Proceedings of the 2017 ACM SIGSAC conference on computer and
  communications security}, pages 363--376, 2017.

\bibitem{zou2019mu}
Deqing Zou, Sujuan Wang, Shouhuai Xu, Zhen Li, and Hai Jin.
\newblock Vuldeepecker: {A} deep learning-based system for multiclass
  vulnerability detection.
\newblock {\em IEEE Transactions on Dependable and Secure Computing},
  18(5):2224--2236, 2019.

\bibitem{zuo2018neural}
Fei Zuo, Xiaopeng Li, Patrick Young, Lannan Luo, Qiang Zeng, and Zhexin Zhang.
\newblock Neural machine translation inspired binary code similarity comparison
  beyond function pairs.
\newblock {\em arXiv preprint arXiv:1808.04706}, 2018.

\end{thebibliography}
